\documentclass[11pt]{article}

\usepackage[final]{acl}

\usepackage{times}
\usepackage{latexsym}

\usepackage[T1]{fontenc}
\usepackage[utf8]{inputenc}

\usepackage{microtype}

\usepackage{inconsolata}

\usepackage{graphicx}

\usepackage{amsmath,amssymb,amsfonts}
\usepackage{subcaption}
\usepackage{multirow}
\usepackage{graphicx}
\usepackage{pifont}
\usepackage{listings} 
\usepackage{xcolor}
\usepackage{colortbl}
\usepackage{multirow}
\usepackage{booktabs}
\usepackage{xcolor}
\usepackage{newtxtext,newtxmath}
\usepackage{xcolor}
\usepackage{mdframed}
\usepackage{kotex}
\mdfsetup{%
linecolor=white,
backgroundcolor=gray!20, 
}
\usepackage{authblk} 
\usepackage{hyperref} 
\usepackage{calligra}
\usepackage[T1]{fontenc}
\usepackage[most]{tcolorbox}
\newtcolorbox{myrqbox}[1]{
  breakable,
  colback=white, % 배경색 (아주 연한 파랑)
  colframe=blue!5!white, % 테두리 색 (진한 파랑)
  fonttitle=\bfseries, % 제목 폰트 (굵게)
  coltitle=black, % 제목 색상 (검정)
  title=#1, % 박스 제목
  arc=2mm, % 모서리 둥글기
  boxsep=2pt, % 텍스트와 테두리 사이 간격
  left=6pt, % 왼쪽 여백
  right=6pt, % 오른쪽 여백
  top=4pt, % 위쪽 여백
  bottom=4pt % 아래쪽 여백
}
\usepackage{algorithm}
\usepackage{algpseudocode}
\usepackage{amsmath}
\usepackage{amsfonts}

\title{\texttt{DITTO}: A Spoofing Attack Framework on Watermarked LLMs via Knowledge Distillation}

\author{Hyeseon An$^1$, Shinwoo Park$^1$, Suyeon Woo$^2$, \and Yo-Sub Han$^{1,\dagger}$ \\
  $^1$Yonsei University, Seoul, Republic of Korea \\
  \texttt{\{\href{mailto:hsan@yonsei.ac.kr}{hsan}, \href{mailto:pshkhh@yonsei.ac.kr}{pshkhh}, \href{mailto:emmous@yonsei.ac.kr}{emmous}\}@yonsei.ac.kr} \\
  $^2$Seoul National University, Republic of Korea \\
  \href{mailto:jenifer7933@snu.ac.kr}{\texttt{jenifer7933@snu.ac.kr}} \\
}

\newcommand{\correspondingfootnote}{
    \let\oldthefootnote=\thefootnote
    \renewcommand{\thefootnote}{}
    \footnotetext{$\dagger$ Corresponding author.}
    \let\thefootnote=\oldthefootnote
}

\begin{document}

\maketitle

\correspondingfootnote 
\begin{abstract}

% Large language model watermarking has been trusted as a core technology for
% proving the authorship of AI-generated content. In order to enhance the robustness
% of watermark algorithm, previous work has focused on advancing detection techniques
% and on countering revomal attempts. However, the security vulnerability inherent in
% the fundamental assumption that
% \textit{a specific watermark can only be generated by its corresponding model}
% has been largely overlooked. We provide the first demonstration that
% the observed phenomenon of watermark \textit{radioactivity} during knowledge
% distillation can be exploited as a spoofing attack against specific model watermarks.
% We propose a novel spoofing attack framework that leverages the watermark radioactivity
% to enable an arbitrary model to mimic a target model's watermark.
% Our findings expose a critical reliability concern in existing watermarking systems
% and strongly suggest the need for next-generation security technologies capable of
% discerning not just the presence, but the authenticity of a watermark.
% Our code is available at \url{https://anonymous.4open.science/r/ditto-8838/}.
The promise of LLM watermarking rests on a core assumption 
that a specific watermark proves authorship by a specific model. 
We demonstrate that this assumption is dangerously flawed. 
We introduce the threat of watermark spoofing, a sophisticated attack that allows 
a malicious model to generate text containing the authentic-looking watermark 
of a trusted, victim model. 
This enables the seamless misattribution of harmful content, 
such as disinformation, to reputable sources. 
The key to our attack is repurposing 
\textit{watermark radioactivity}, the unintended inheritance of data patterns 
during fine-tuning, from a discoverable trait into an attack vector. 
By distilling knowledge from a watermarked teacher model, 
our framework allows an attacker to steal and replicate the 
watermarking signal of the victim model. 
This work reveals a critical security gap in text authorship verification 
and calls for a paradigm shift towards technologies capable of 
distinguishing authentic watermarks from expertly imitated  ones.
Our code is available at \url{https://github.com/hsannn/ditto}.

\end{abstract} 

\section{Introduction}
\label{sec:introduction}
Large language models~(LLMs) are rapidly permeating nearly every aspect 
of modern life,
from industrial applications and 
education to 
everyday personal use~\citep{goyal2022news,xiao2024humanaicollaborativeessayscoring}.
Their ability to generate coherent, 
human-like text at scale makes them powerful tools for
productivity and innovation, but also raises concerns about 
authenticity and trust~\citep{perkins2023academic,wang2024megafake}.
As these models become increasingly integrated into daily workflows and 
public discourse, the ability to detect and verify LLM-generated text has become 
ever more crucial~\citep{mireshghallah-etal-2024-smaller,go-etal-2025-xdac,park-etal-2025-katfishnet}.

As regulators in the United States \citep{biden2023executive} 
and the EU AI Act \citep{euaiact2024}
have signalled the need for clearer provenance of LLM-generated content, it becomes critical
not only to determine whether a piece of text was authored by a human or generated by an AI system
but also to identify which model generated it.
Notably, major industry players, such as Meta, OpenAI, and Google Deepmind, 
have focused on watermarking as
a practical tool for provenance verification \citep{reuters2023watermark,dathathri2024scalable}.
This provenance is essential for accountability, since different models may be governed by distinct
licenses, exhibit varying reliability, or present different risks of misuse.
For these reasons, the development of 
LLM watermarking~\citep{kirchenbauer2023watermark,kuditipudi2024robust,lee-etal-2024-wrote,gloaguen2025watermarking,park2026watermod,kim2026marking} 
techniques has attracted substantial interest.

\begin{figure}[t!]
    \centering
    \includegraphics[width=\linewidth]{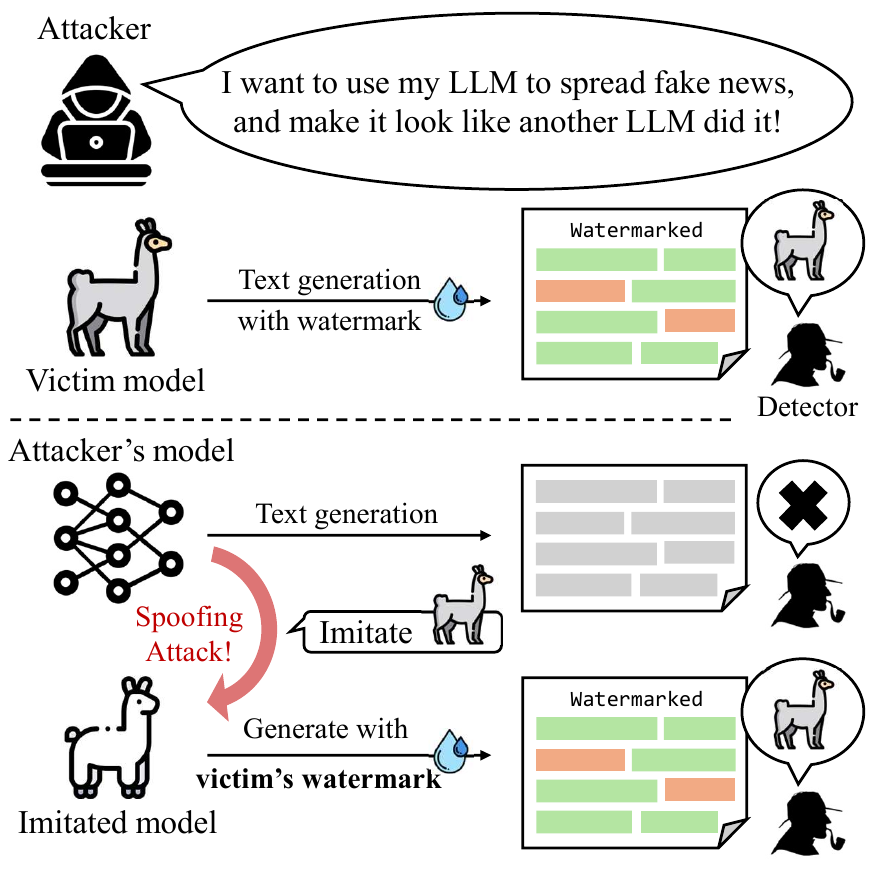}
    \caption{
    % We `do something bad' to reveal the fragility of current LLM watermarking
    % schemes. By imitating the watermark of the victim LLM, we can identify breaking points
    % before real adversaries do.
    We `do something bad' to reveal the fragility of current LLM watermarking schemes. 
    An attacker’s model is trained to imitate the watermark of a 
    trusted victim model, enabling it to generate content~(e.g., fake news) 
    that is then falsely attributed to the victim by a detector.
    }
    \label{fig:threat_scenario}
\end{figure}

LLM watermarking---pioneered in earnest by \citet{kirchenbauer2023watermark}---refers to
algorithms that embed an imperceptible but
machine-detectable signal into model outputs so as to convey authorship information.
In practice, watermarking has been explored through two main paradigms: logit-bias schemes
that partition the vocabulary and skew generation probabilities 
for detectability~\citep{kirchenbauer2023watermark,lu-etal-2024-entropy,chen-etal-2024-watme,guo2024context},
and sampling-based methods that modify the decoding procedure to embed verifiable signals
with minimal impact on text quality~\citep{kuditipudi2024robust,dathathri2024scalable}.

Despite its promise, LLM watermarking has not been free from criticism, and a growing
body of work has examined the limitations and shortcomings of existing schemes
through a variety of attacks. 
Paraphrasing-based attacks~\citep{rastogi-pruthi-2024-revisiting,cheng2025revealing} 
revealed straightforward weaknesses, 
while more sophisticated techniques have probed watermark robustness 
more deeply~\citep{wu-chandrasekaran-2024-bypassing,jovanovic2024watermark,chen2024mark}.
By demonstrating how watermarks can be removed or reverse-engineered,
these studies highlight the practical limitations of current schemes and motivate
the development of stronger defenses.

Notably, most prior attack work has focused on \textit{scrubbing attacks} that aim to remove 
embedded signals to evade detection.
Relatively less attention has been given to \textit{spoofing attacks},
which aim to make a third-party model generate outputs that contain the detectable
watermark patterns of a target model, thereby misleading attribution.
Spoofing attack presents qualitatively distinct and more malicious threat.
Unlike scrubbing, it fabricates the watermark, leading detectors to falsely 
attribute text to the wrong model. Such misattribution produces a misleading
sense of certainty, which is especially hazardous in the spread of
fake news, political propaganda and other forms of disinformation.
Therefore, a comprehensive assessment of watermark robustness should consider
not only whether a watermark can be erased, but also how easily it can be forged.

We address this underexplored threat by proposing an attack framework for 
spoofing LLM watermarks under a practical black-box setting.
Figure~\ref{fig:threat_scenario} illustrates our attack scenario:
an adversarial user, seeking to spread disinformation such as fake news,
may prefer that the text appear to originate from a trusted system like ChatGPT
rather than from their own local model.
The attacker has neither the logits of the target model nor knowledge of its watermarking scheme.
Under these realistic constraints, 
our approach exploits
\textit{watermark radioactivity}~\citep{sander2024watermarking,sander2025detecting}---the phenomenon whereby
training on synthetic outputs generated by a watermarked model causes 
downstream models to inherit the same watermark signal.

We introduce \textbf{\texttt{DITTO}}~(\textbf{D}istilled watermark \textbf{I}mitation of
a \textbf{T}argeted \textbf{T}eacher's \textbf{O}utputs),
a knowledge-distillation-based spoofing attack framework.

Through our approach, 
we show that watermarking is vulnerable not only to scrubbing attacks
but also to a more fundamental threat, spoofing attacks.
Consequently, our work highlights the need for security mechanisms that go beyond
mere presence detection to include authenticity verification.

We summarize our contribution as follows:
\begin{itemize}
    % \item We are the first to exploit the watermark radioactivity for spoofing,
    % repurposing the signal from a detection mechanism into an attack vector.
    \item We are the first to repurpose watermark radioactivity for spoofing attacks, fundamentally repurposing a phenomenon intended for detection into a potent tool for misattribution.
    \item We show that our attack is highly adaptable, proving effective against not only n-gram-based watermarks but also sampling-based schemes.
    \item We find that the attack breaks the conventional trade-off between strength and quality.
    The spoofing intensity can be significantly increased without a discernible degradation in text quality,
    making the attack highly evasive.
\end{itemize}

\section{Related Work}
\label{sec:related_work}
The popularization of LLMs highlights the necessity of robust safeguards for their
responsible use. Similar to traditional cybersecurity, the field of LLM watermarking
has co-evolved through a dynamic interplay between the development of new schemes
and the attack systems that test their robustness.

\subsection{LLM Watermarking}
Early work on LLM watermarking centers on the green-red list paradigm, which partitions
the vocabulary and biases sampling toward a pseudorandom `green' set to yield an imperceptible
yet statistically detectable signal \citep{kirchenbauer2023watermark}.
Follow-ups adapt the idea to demanding settings: \citet{lee-etal-2024-wrote} selectively
applies the green-red bias only when token entropy is high, markedly improving detectability
and preserving code correctness under the low-entropy dynamics of code generation;
multi-bit schemes add position allocation for user-traceable identifiers \citep{yoo-etal-2023-robust};
entropy-guided insertion further tunes watermark timing for robustness \citep{cai2025entropy};
and context-aware methods build semantically balanced green-red lists to resist paraphrasing
while limiting quality loss \citep{guo2024context}.
Complementary detection advances weight tokens by entropy to boost sensitivity in low-entropy
regimes \citep{lu-etal-2024-entropy}.
Collectively, these works expand list-based watermarking from a single detection signal
toward more accountable and semantically informed mechanisms.

In parallel, sampling-based watermarking modifies the sampling rule itself rather than fixing
token partitions. \citet{dathathri2024scalable} introduced tournament sampling with configurable
non-distortion guarantees and showed production-scale viability and improved detectability at a
fixed false-positive rate.
Distortion-free approach, such as \citet{kuditipudi2024robust}, alter the sampling process
to embed watermarks while keeping the model’s output distribution unchanged, ensuring
detectability even after edits.
Together these lines trace a shift from fixed list biases to selective, semantic
and sampling-theoretic designs that better balance robustness, attribution, and text quality.

\subsection{Watermark Attack}
As watermarking techniques for LLMs have been developed, so too have methods to attack them.
Recent research on watermark attacks is evolving beyond simple text modification towards
stealing techniques \citep{wu-chandrasekaran-2024-bypassing,jovanovic2024watermark,chen2024mark} that first uncover the watermark's mechanism.
This act of stealing enables more sophisticated \textbf{scrubbing} and \textbf{spoofing} attacks.
Scrubbing attack aims to remove the watermark to evade detection and
spoofing attack causes unwatermarked text to be accepted as watermarked.
\citet{jovanovic2024watermark} showed that black-box querying suffices to learn the
watermarking rule well enough, calling into question optimistic deployment claims.
Beyond heuristic estimation, \citet{zhang2024large} claimed that mixed-integer programming can recover
KGW-style green lists and remove the watermark without detector access or watermark metadata.

A fast-growing line of research has focused on watermark scrubbing under realistic constraints.
One major approach is \citet{wu2024bypassing}, which infers the watermark's token `colors'
to strategically replace green tokens with non-green ones, while
\citet{huang-etal-2025-b4} framed scrubbing as constrained optimization and distilled a proxy
distribution to enable strong black-box watermark removal.
Recently, \citet{chen2024mark} estimated n-gram parameters (prefix length, strength, red/green lists)
and provided theoretical bounds on distributional closeness after removal.
Complementing removal, \citet{pang2024no} revealed that common design choices intended
to make LLM watermarks robust can make the systems vulnerable to simple attacks, exposing a
fundamental trade-off. However, prior work has concentrated on stealing and scrubbing,
with spoofing often treated as a derivative capability or explored only in specific contexts.
Our study, therefore, adopts a spoofing-first perspective to close this gap.

\subsection{Watermark Radioactivity}
Watermark radioactivity is the unintended inheritance of watermark signals when a student
LLM is fine-tuned on watermarked teacher outputs.
Even small fractions of watermarked data can make radioactivity statistically detectable,
offering a tool for provenance auditing \citep{sander2024watermarking}.
However, \citet{pan-etal-2025-llm} demonstrated that adversaries can largely erase these signals through
paraphrasing or inference-time neutralization.
Building on this line of work, we leverage watermark radioactivity as a foundation
for developing a scheme-agnostic spoofing methodology.

\begin{figure*}[h!]
    \centering
    \includegraphics[width=\textwidth]{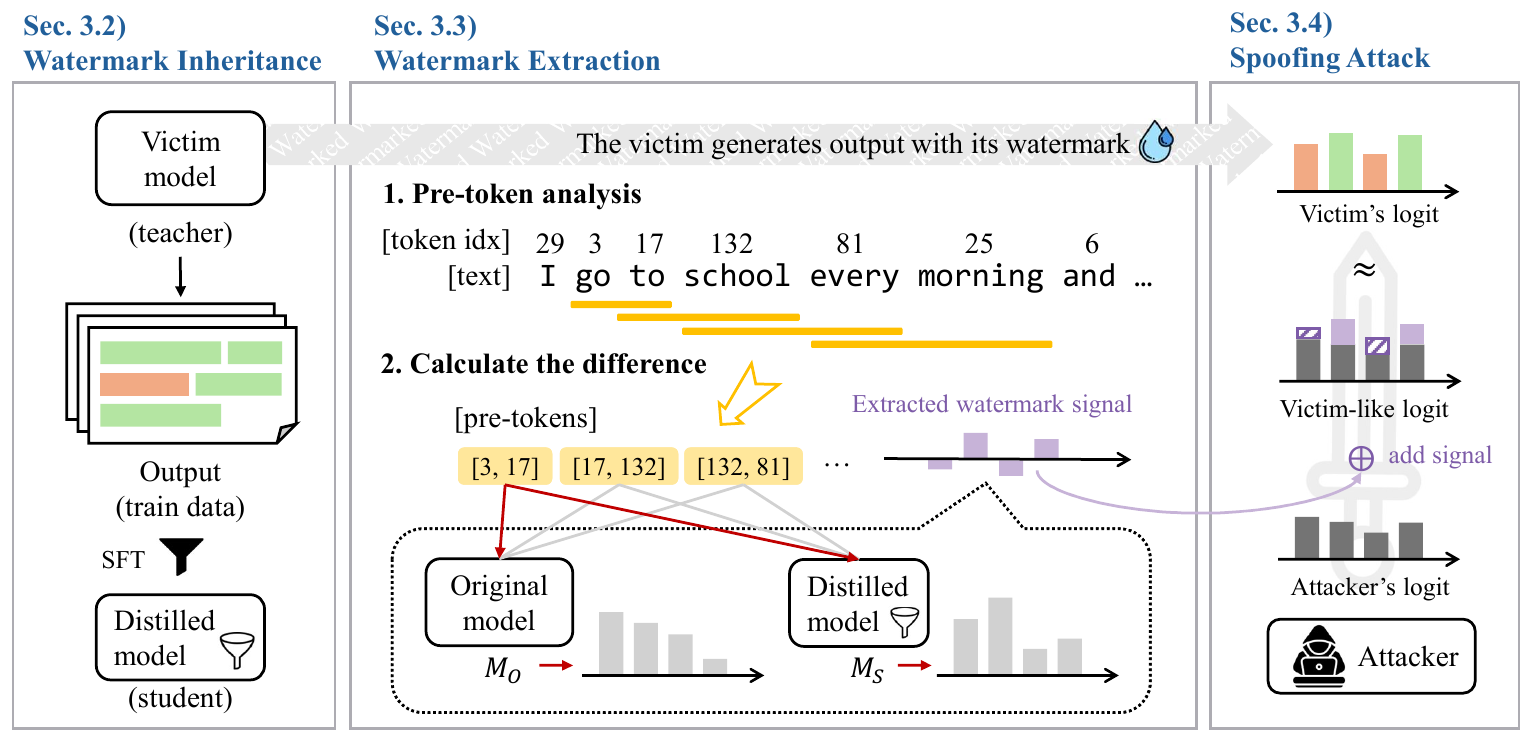}
    \caption{An overview of the \texttt{DITTO} framework. Our method consists of three main stages:
    (1)~Watermark Inheritance, where a student model learns the teacher's watermarked patterns via knowledge distillation;
    (2)~Watermark Extraction, where the watermark is isolated by analyzing averaged logit differences, both globally and for specific text prefixes;
    and (3)~the Spoofing Attack, where the extracted signal is added to the attacker's logits to imitate the victim's watermark.}
    \label{fig:overview}
\end{figure*}

\section{DITTO}
\label{sec:approach}
We propose \texttt{DITTO}, a spoofing attack framework that leverages the radioactivity of LLM watermark,
which is the tendency for a model to inherit statistical patterns when trained on the outputs
of a watermarked source.
The objective of \texttt{DITTO} is to imitate the unique watermark signal of a specific target model $M_T$,
enabling an arbitrary external model $M$ to generate text that appears as if it were
watermarked by the target. Figure~\ref{fig:overview} demonstrates \texttt{DITTO} framework.
First, we use knowledge distillation to transfer the watermark from $M_T$
into a student model $M_S$. Then, we extract the watermark pattern by analyzing the difference
in output distributions between $M_S$ and its original version $M_O$.
Lastly, we inject this extracted signal directly into the logits of $M$,
successfully replicating the target's watermark.

\subsection{Preliminaries}
Let $M$ denote a large language model with a vocabulary $V$.
Given a prefix (or context) $c = (x_1, ..., x_{t-1})$, the model generates a
probability distribution $P_M(x_t|c)$ for the next token $x_{t}$.
This $P_M(x_t|c)$ is typically derived by applying a softmax function to the
$M$'s output logits $l_M(c) \in \mathbb{R}^{|V|}$.

Our attack aims to imitate a specific watermarked model, which we refer to as the target model $M_T$.
This model embeds a statistical signal into its output by systematically altering its next-token
probability distribution. Specifically, the watermarking process modifies the original logits
with a context-dependent signal vector, $\delta(c)$.
The original logits $l_{M_T}(c)$ are adjusted to produce the watermarked logits $l'_{M_T}(c)$:
\begin{equation}
    l'_{M_T}(c) = l_{M_T}(c) + \delta(c).
\end{equation}
Consequently, the final watermarked probability distribution is given by:

\begin{equation}
    P'_{M_T}(x_t|c) = \text{softmax}(l_{M_T}(c) + \delta(c)).
\end{equation}

This modification subtly increases the likelihood of selecting certain tokens based on the prefix,
thereby embedding a detectable pattern in the generated text without significantly degrading its quality.
Our goal is to reverse-engineer an approximation $\delta(c)$ to imitate $M_T$'s watermark.

\subsection{Watermark Inheritance}
The initial step of \texttt{DITTO} is to transfer the watermark pattern from the target model $M_T$
to an open-source model whose output logits are accessible.
We accomplish this through knowledge distillation, which facilitates the transfer
of the target model's watermark signal to a student model.

We first generate a large text dataset $D_T$ using the watermarked target model $M_T$.
Every text in $D_T$ is imperceptibly embedded with the unique watermark pattern of $M_T$.
We then train the original student model $M_O$, on this data to obtain the trained student
model $M_S$ using a standard supervised fine-tuning (SFT) objective.

\begin{equation}
    \theta_{S} = \underset{\theta_O}{\arg\max} \sum_{\mathbf{x} \in D_T} \sum_{i=1}^{|\mathbf{x}|} \log P(x_i | x_{1:i-1}; \theta_O),
\end{equation}
where $\theta_{S}$ and $\theta_{O}$ are denote the model parameters for $M_S$ and $M_O$ respectively
and $\mathbf{x}$ is a text sequence from the training dataset $D_T$.

As a result, the output distribution of the trained student model, $P_{M_S}$, becomes
a close approximation of the target model's watermarked distribution $P'_{M_T}$ for contexts
seen during training:

\begin{equation}
    P_{M_{S}} (x_t | c) \approx P'_{M_T} (x_t | c), \quad \forall c \in D_T.
\end{equation}

Due to the watermark radioactivity, the outputs of $M_S$ exhibit similar watermarking patterns
with $M_T$. This provides the foundation for extracting the watermark signal in the next step.

\subsection{Watermark Extraction}
\label{subsec:ews}
The goal of this stage is to extract the watermark pattern embedded within the trained
student model $M_S$ as a quantitative signal, which we term
the Extracted Watermark Signal (EWS). In order to obtain the signal required
for our spoofing attack, we adapt the watermark stealing methodology from \citet{pan-etal-2025-llm}
as an effective tool.

The core principle is to isolate the systematic bias induced by the watermark
by comparing the behavior of the original model before training ($M_O$) and the model
after training ($M_S$).
First, the context-independent global bias $\delta_{\text{global}}$ is calculated
as the average difference in logits between the two models over the entire training dataset:

\begin{equation}
    \delta_{\text{global}} = \mathbb{E}_{c \in D_T} [l_{M_{S'}}(c)] - \mathbb{E}_{c \in D_T} [l_{M_O}(c)].
\end{equation}

Similarly, the local bias $\delta_p$ for a specific $n$-gram prefix $p$ is calculated by
narrowing the scope to contexts ending with that prefix:

\begin{align}
    \delta_p = & \;\mathbb{E}_{c \in D_T | c \text{ ends with } p} [l_{M_{S}}(c)] \\
& \;\; - \mathbb{E}_{c \in D_T | c \text{ ends with } p} [l_{M_O}(c)]. \nonumber
\end{align}

These averaging processes are
designed to cancel out context-specific knowledge shifts,
leaving only the consistent watermark pattern.

The final EWS is constructed as a weighted sum of these two biases:
\begin{equation}
    \text{EWS}(c) = \delta_{\text{global}} + \sum_{p \in \text{prefixes}(c)} w(p) \cdot \delta_p.
\end{equation}
Prefixes with a higher frequency in the training data are assigned
greater weights, reflecting their stronger influence.
Following \citet{pan-etal-2025-llm}, to ensure prefixes with stronger statistical signals have a
greater impact, we assign a weight $w(p)$ to each prefix's local bias. This weight
is determined by the frequency of the prefix $p$ in the training data (\textit{i.e.},
the output of teacher model) giving greater influence to more common prefixes.

This extracted watermark signal quantitatively replicates the target model's
watermark and serves as the key ingredient for the spoofing attack we
propose in the following section.

\subsection{Spoofing Attack}
Using the watermark signal $\text{EWS}(c)$ extracted in the previous stage,
we perform a spoofing attack to replicate the watermark of the target model $M_T$ onto
an unwatermarked model.

The attack is executed in real-time during the inference phase of text generation.
Just before the unwatermarked original model $M_O$ enerates the next token,
the corresponding $\text{EWS}(c)$ vector for the current context $c$ is retrieved.
The EWS vector is then added to the model's original logit vector $l_{M_O}(c)$,
to yield the final modified logits $l'_{M_O}(c)$:

\begin{equation}\label{eq:spoof}
    l'_{M_O}(c) = l_{M_O}(c) + \alpha \cdot \text{EWS}(c),
\end{equation}
where $\alpha$ is a scaling parameter that controls the injection strength.
The impact of the scaling parameter α on the performance of the spoofing attack is discussed further in Section~\ref{sec:discussion}.

The modified logits guide the model to follow the statistical bias of the target model
during the next-token sampling process. Consequently, the attacked $M_O$ generates
text embedded with the target model's unique watermark, thereby successfully
completing the spoofing attack.

\section{Results of \texttt{DITTO}}
\label{sec:experimental_results}

% 프리앰블
% \usepackage{booktabs}
% \usepackage{multirow}

\subsection{Experimental Setup}
\label{subsec:setup}
In order to validate the effectiveness of \texttt{DITTO}, we design the experimental environment as follows.

\paragraph{Models.}
we utilize several LLMs that are widely used in recent research: Llama3.1-8B~\footnote{\url{https://huggingface.co/meta-llama/Llama-3.1-8B-Instruct}},
Llama3.2-3B~\footnote{\url{https://huggingface.co/meta-llama/Llama-3.2-3B-Instruct}} and
Llama3.2-1B~\footnote{\url{https://huggingface.co/meta-llama/Llama-3.2-1B-Instruct}}.
These three models are paired as teacher and student models for our experiments.
The specific pairings are following Table~\ref{tab:model} if not specified.
The attack model shares the same backbone architecture as the corresponding student model. 
For example, to imitate the watermark of a Llama3.1-8B teacher,
a Llama3.2-3B model first extracts the watermark signal as the student,
and a distinct Llama3.2-3B then serves as the attack model to perform the spoof.
We employ Llama2-14B to measure the perplexity of the generated text.
\begin{table}[h!]
    \centering
    \begin{tabular}{c|c}
        \toprule
        \textbf{Teacher Model} & \textbf{Student Model} \\ 
        \midrule
        Llama3.1 8B Instruct & Llama3.2 3B Instruct \\ 
        Llama3.2 3B Instruct & Llama3.2 1B Instruct \\ 
        % Qwen3 4B    & Qwen2.5 7B \\ 
        \bottomrule
    \end{tabular}
    \caption{Teacher-student model pairs used in the knowledge distillation.
    Each student inherits watermarking signal from its corresponding teacher model.}
    \label{tab:model}
\end{table}

\paragraph{Datasets.}
For model training and evaluation, we use two distinct datasets.
Dolly-15k \citep{conover2023free} is an open-source dataset of 15,000 human-generated prompt/response
pairs, to fine-tune the models for interactive, instruction-based capabilities.
To specifically assess watermarking performance on structured and narrative text, we also use
the Bookreport task from the MarkMyWords (MMW) benchmark \citep{piet2025markmywords},
a framework designed for the systematic evaluation of watermarking techniques.

\paragraph{Training and Attack Parameters.}
For student model training, we do Supervised Fine-Tuning (SFT) with Low-Rank Adaptation (LoRA)
and train for a total of 3 epochs. From the total dataset of 10,000 samples,
we hold out 100 samples as a test set for evaluation.
Watermark injection and detection were based on the algorithm by \citet{kirchenbauer2023watermark},
with parameters set to $\delta=3$ and $\gamma=0.5$ for injection, and a $z$-threshold of 4.0 for detection.
For spoofing attack, $\alpha$, which controls the strength of the spoofing attack,
is varied across the range of [2.5, 3, 3.5, 4, 4.5, 5].
Unless otherwise specified, all main results reported in our tables use a fixed $\alpha$ of 4.5.

\subsection{Evaluation Metrics}
\paragraph{\textbf{TPR@FPR.}}
The True Positive Rate (TPR) measures how often a watermark is correctly found at a
fixed False Positive Rate (FPR). To create a strict test, we set the FPR to low levels (10\%, 1\%, and 0.1\%),
then measure the resulting TPR for our spoofed texts.

\paragraph{p-value.}
The statistical strength of a detected watermark is measured by its p-value,
where a lower value indicates stronger evidence. A strict threshold is required to prevent
false alarms when analyzing long texts. We report the median p-value to show the overall
strength of the spoofed signal.

\subsection{Experimental Results}
\label{subsec:results}
% Table~\ref{tab:main} presents the results of our watermark spoofing attack across various model pairings.
The performance of our proposed methodology is evaluated against several baselines.
First, we establish the original watermarked model as a baseline to define the upper bound for the
watermark detection rate. Additionally, we include prior works on spoofing attacks,
JSV \citep{jovanovic2024watermark} and De-Mark \citep{chen2024mark}, as primary comparison targets.
Notably, De-Mark considers two settings based on the attacker's level of access.
The graybox setting assumes an ideal scenario where the attacker has direct access to the model's
top-k token probabilities to infer the watermarking rule.
In contrast, the blackbox setting represents a more realistic and challenging attack environment where,
lacking probability information, the attacker must indirectly deduce the rule from the frequency
of multiple generation outputs. Since our methodology operates under a blackbox assumption,
a fair comparison is made against De-Mark (blackbox).

The results of our spoofing attack on Llama3.1-8B and Llama3.2-3B are presented in Table~\ref{tab:main},
alongside the baselines. The results demonstrate that \texttt{DITTO} can successfully deceive
watermark detectors by injecting imitated watermark signal into the student model's output.
Notably, our methodology proves to be more effective when the backbone model size is smaller.
We provide a detailed analysis of these results in the subsequent sections.

\begin{table*}[h!]
    \centering\resizebox{\textwidth}{!}{
    \small
    \begin{tabular}{l|ccccc|ccccc}
        \toprule
        & \multicolumn{5}{c|}{\textbf{MMW Bookreport}} & \multicolumn{5}{c}{\textbf{Dolly CW}} \\
        \cmidrule(lr){2-11}
        & \multicolumn{3}{c}{TPR@FPR=} 
        & \multirow{2}{*}[-2pt]{\shortstack{median\\p-value $\downarrow$}} & \multirow{2}{*}{ppl $\downarrow$} 
        & \multicolumn{3}{c}{TPR@FPR=} 
        & \multirow{2}{*}[-2pt]{\shortstack{median\\p-value $\downarrow$}} & \multirow{2}{*}{ppl $\downarrow$}  \\
        \cmidrule(lr){2-4} \cmidrule(lr){7-9}
         &  10\%$\uparrow$ & 1\%$\uparrow$ & 0.1\%$\uparrow$ & & & 10\%$\uparrow$ & 1\%$\uparrow$ & 0.1\%$\uparrow$ & & \\
        \midrule
        Llama3.1-8b Watermarked &  0.89  & 0.75  & 0.65  & 1.10E-06  & 3.87 & 0.92 & 0.82  & 0.70  & 1.88E-07  & 6.34 \\
        \quad + JSV            &  0.36  &  0.10 & 0.04  & 2.20E-01  & \textbf{2.81}  & 0.56 & 0.26  & 0.11  & 6.00E-02  & 6.62     \\
        \quad + De-Mark (gray) & 0.99  & \textbf{0.96}  & \textbf{0.88}  & \textbf{5.73E-06}  & 3.77  & 0.96 & 0.89  & 0.77  & 4.32E-05  & 3.97   \\
        \quad + De-Mark (black)&  \textbf{1.00}  & 0.94  & 0.73  & 6.93E-05  & 3.55  & \textbf{0.97} & 0.82  & 0.70  & 2.66E-05  & 3.93       \\
        \quad \textbf{+ \texttt{DITTO}~(Ours)}  &  0.81  & 0.70  & 0.51  & 7.97E-04  & 4.18  & 0.96 & \textbf{0.93}  & \textbf{0.85}  & \textbf{1.78E-07}  &  \textbf{3.73}     \\
        \midrule
        Llama3.2-3b Watermarked &   0.91   & 0.79  & 0.74  & 3.43E-07  & 4.28  & 0.82 & 0.68  & 0.61  & 2.64E-06  & 6.34  \\
        \quad + JSV            &   0.70   & 0.34  & 0.18  & 4.00E-02  & 3.78  & 0.73 & 0.42  & 0.28  & 1.30E-02  & 6.93      \\
        \quad + De-Mark (gray) &   0.98   & 0.96  & 0.90  & 1.93E-06  & 4.65  & 0.99 & \textbf{0.93}  & 0.75  & 2.66E-05  &   4.64    \\
        \quad + De-Mark (black)& \textbf{1.00} & 0.96  & 0.90  & 3.34E-06  & 4.68  & \textbf{1.00} & 0.90  & \textbf{0.80}  & 2.66E-05  & 4.74      \\
        \quad \textbf{+ \texttt{DITTO}~(Ours)}  & 0.99 & \textbf{0.99}  & \textbf{0.97}  & \textbf{5.48E-17}  & \textbf{2.44}  & 0.91 & 0.86  & 0.75  &  \textbf{1.78E-07} & \textbf{2.95} \\
        \bottomrule
    \end{tabular}}
    \caption{Spoofing-attack evaluation results on MMW Book Report and Dolly Creative Writing datasets. We report detection performance (TPR at fixed FPRs), median p-values and perplexity (ppl) for different watermarking defenses.
    Arrows indicate the preferred direction for each metric ($\uparrow$: higher is better, $\downarrow$: lower is better).}
    \label{tab:main}
\end{table*}

\section{Analysis}
\label{sec:discussion}
Section~\ref{sec:experimental_results} demonstrates that \texttt{DITTO} can successfully mimic the target model's
watermark signal. This motivates a deeper exploration into the conditions that maximize effectiveness,
as well as an investigation of its limitations and potential for expansion.
% Accordingly, this section provides a multifaceted analysis of \texttt{DITTO}'s characteristics by
% proposing and investigating the following three research questions.

\subsection{Impact of Spoofing Strength}
\label{subsec:alpha_scaling}

A successful spoofing attack must not only achieve a high detection rate but also maintain \textit{stealthiness}
by preserving the original text quality to avoid suspicion. The key variable that balances these two objectives
is the scaling parameter $\alpha$ (Eq.~\ref{eq:spoof}).
Here, we conduct an in-depth analysis of how varying $\alpha$ affects the spoofing strength
and the quality of the generated text to determine the efficiency and limitations of our proposed methodology.
We experiment with the same scenarios as Section~\ref{sec:experimental_results},
imitating Llama3.1-8B with Llama3.2-3B and Llama3.2-3B with Llama3.2-1B,
increasing $\alpha$ from 2.5 to 5.0 in increments of 0.5.

\begin{figure}[t!]
    \centering
    \begin{subfigure}[b]{\columnwidth}
        \includegraphics[width=\columnwidth]{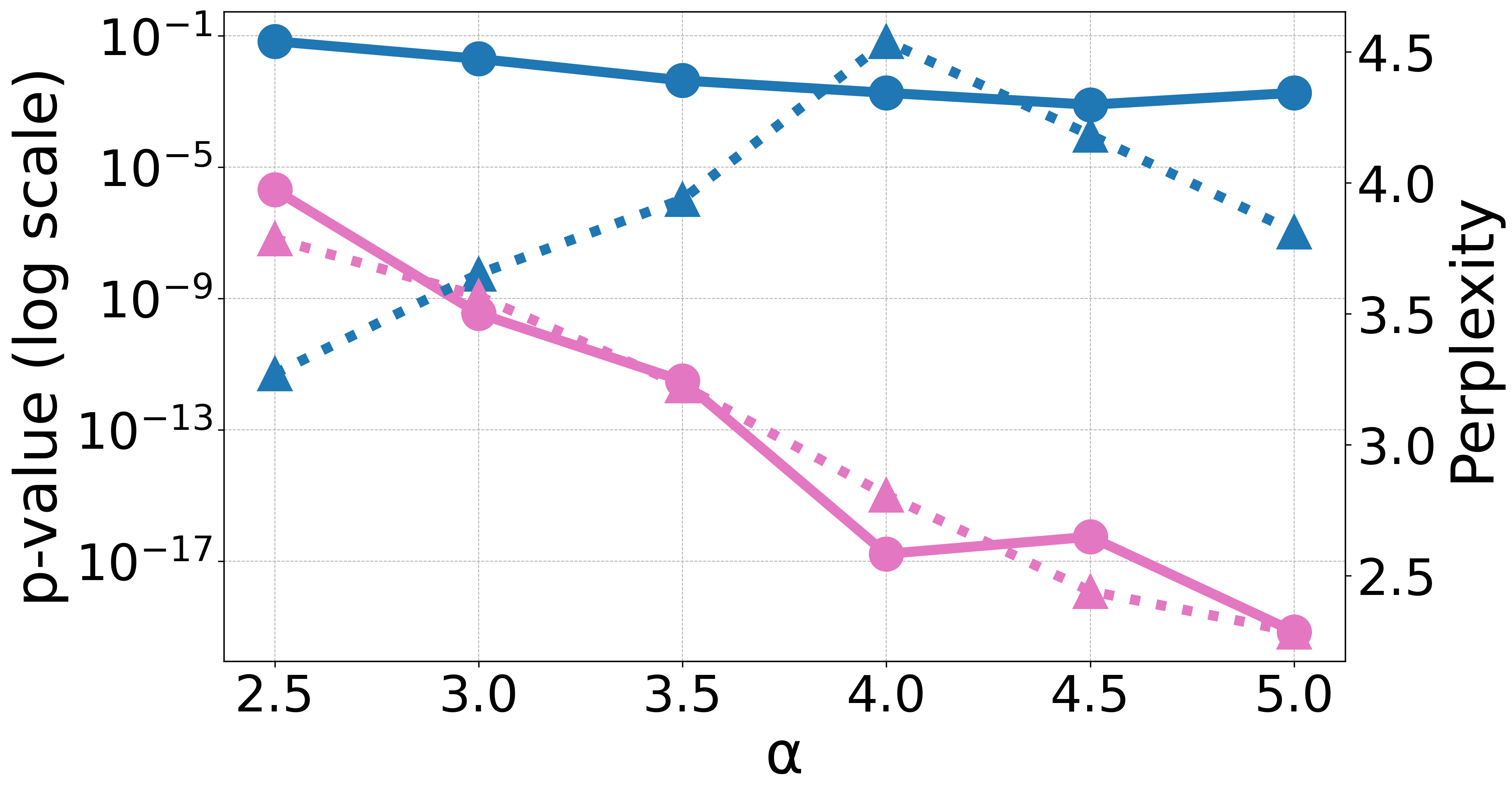}
        \caption{MMW Bookreport}
        \label{fig:delta_scale_mmw}
    \end{subfigure}
    
    \begin{subfigure}[b]{\columnwidth}
        \includegraphics[width=\columnwidth]{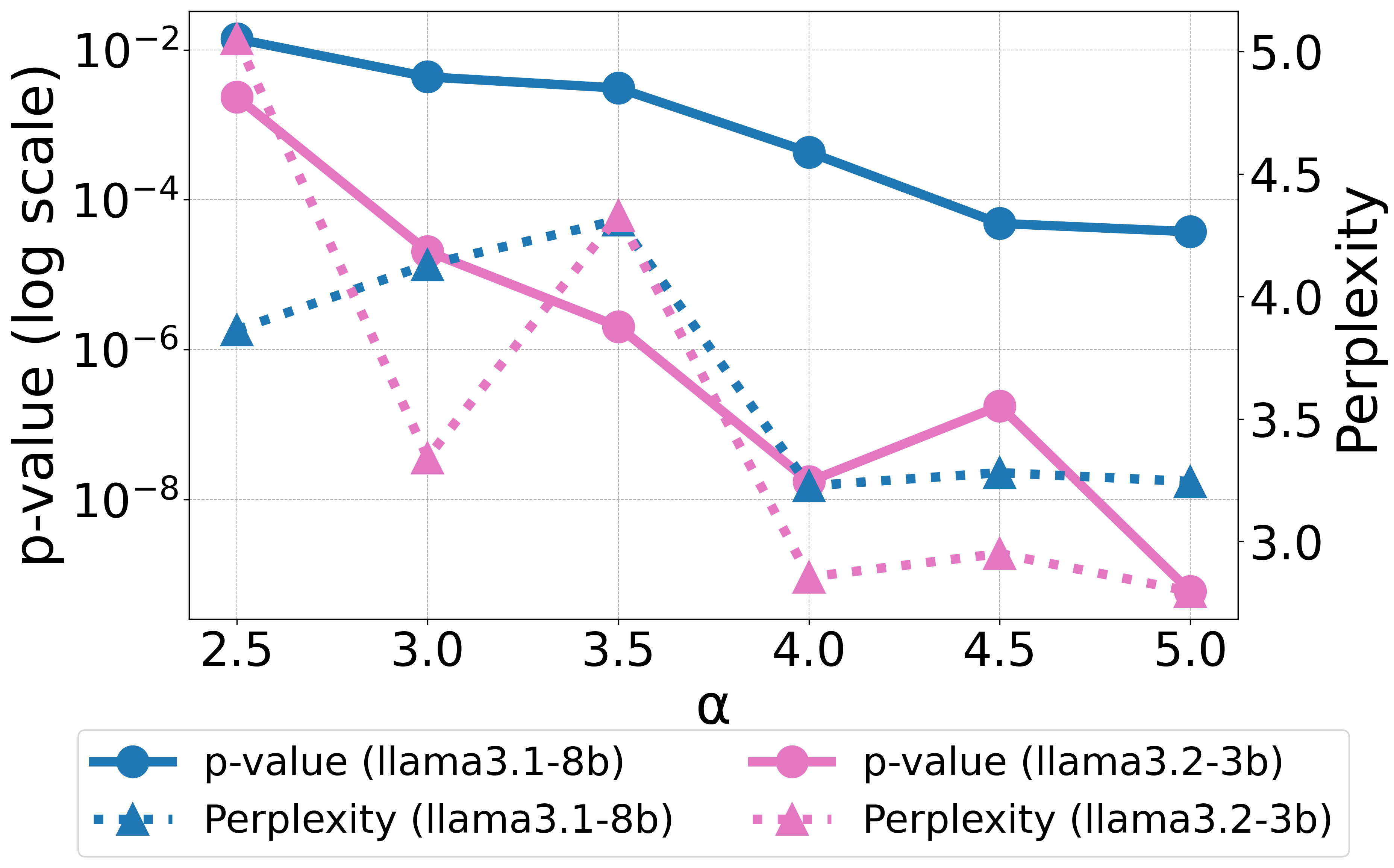}
        \caption{Dolly CW}
        \label{fig:delta_scale_dollycw}
    \end{subfigure}
    \caption{The effect of varying the $\alpha$ parameter on the p-value and Perplexity.
    The complete experimental results for this analysis are available in Table~\ref{tab:alpha_full} in Appendix~\ref{appendix:full}.}
    \label{fig:alpha}
\end{figure}

We initially hypothesize that a larger $\alpha$ would lead to a higher spoofing success rate at the
cost of degraded text quality (\textit{e.g.}, higher perplexity), creating a clear trade-off.
However, the experimental results presented in Figure~\ref{fig:alpha} reveal a surprising trend
that deviates from our expectation. 
As expected, across both the MMW Bookreport~(\ref{fig:delta_scale_mmw})
and Dolly CW~(\ref{fig:delta_scale_dollycw}), the median p-value exhibits a decrease as $\alpha$ grows.
However, the perplexity defies our prediction.
Instead of a steady increase, the text quality metric exhibits a non-monotonic pattern;
while it rises in some intervals, it frequently decreases in others as the attack intensifies.
This volatility is particularly pronounced when imitating the Llama3.2-3b model,
a result that challenges the common assumption that stronger attacks must necessarily degrade text quality.

This finding reveals a critical characteristic of the \texttt{DITTO} attack:
an adversary can modulate the spoofing strength without producing a signal of quality degradation.
Detection methods that operate on the heuristic that attacks degrade text quality
are likely to fail, as our results show that perplexity fluctuates unpredictably
rather than consistently increasing with spoofing strength.
This finding, therefore, implies that \texttt{DITTO} possesses a greater potential for stealth
and evasion than a simple trade-off relationship would suggest.

% \begin{myrqbox}{RQ2}
% Does the attack work by targeting a flaw in a specific watermarking method,
% or by learning the general statistical signal that all watermarks leave behind?
% \end{myrqbox}

\subsection{Beyond Green-Lists}
While the previous sections demonstrate spoofing capabilities of \texttt{DITTO} against KGW, an n-gram-based green-list watermark,
a critical question remains:
% \begin{quote}
%     \textit{``Does our attack mechanism merely exploit scheme-specific features of watermark,
%     or does it learn the more fundamental patterns of distributional distortion caused by any watermark?''}
% \end{quote}
\begin{mdframed}
    Does our attack mechanism merely exploit scheme-specific features of watermark,
    or does it learn the more fundamental patterns of distributional distortion caused by any watermark?
\end{mdframed}

Answering this is crucial for evaluating the generalizability of our methodology.
In order to investigate this, we test \texttt{DITTO} against SynthID \citep{dathathri2024scalable},
a prominent sampling-based watermark that operates differently from that of \citet{kirchenbauer2023watermark}.
The experiment is conducted on the Dolly CW dataset,
on which our main experiments show the most stable performance.

\begin{table}[h!]
    \centering\resizebox{\columnwidth}{!}{
    \small
    \begin{tabular}{l|ccccc}
        \toprule
        & \multicolumn{3}{c}{TPR@FPR=} 
        & \multirow{2}{*}[-2pt]{\shortstack{median\\p-value $\downarrow$}} & \multirow{2}{*}{ppl $\downarrow$} \\
        &  10\%$\uparrow$ & 1\%$\uparrow$ & 0.1\%$\uparrow$ \\
        \midrule
        \multicolumn{6}{c}{\textbf{Llama3.1-8b (victim)}}  \\
        \midrule
        Watermarked & 0.98 & 0.98 & 0.96 & 1.57E-12 & 2.80 \\
        \textbf{\texttt{DITTO}} & 0.88 & 0.84 & 0.80 & 7.10E-10 & \textbf{2.79} \\
        \midrule
        \multicolumn{6}{c}{\textbf{Llama3.2-3b (victim)}}  \\
        \midrule
        Watermarked & 0.97 & 0.97 & 0.94 & 5.51E-15 & 6.30   \\
        \textbf{\texttt{DITTO}} & 0.90 & 0.89 & 0.87 & 8.12E-12 & \textbf{2.88} \\
        \bottomrule
    \end{tabular}}
    \caption{Spoofing attack results against the SynthID watermark on the Dolly CW dataset.
    \texttt{DITTO} achieves high TPR and low p-values, comparable to the authentically watermarked models.}
    \label{tab:synthid}
\end{table}

Table~\ref{tab:synthid} clearly demonstrates that \texttt{DITTO} achieves highly
effective spoofing performance against SynthID as well.
The text generated by \texttt{DITTO} records high TPRs and low p-values, comparable to the original
watermarked model, successfully deceiving a detector into recognizing it as authentic watermarked text.
The perplexity results are particularly noteworthy. For Llama3.1-8b, \texttt{DITTO}'s perplexity is nearly identical to the original's.
For Llama3.2-3b, our perplexity is substantially lower than that of the original
watermarked model.
This suggests that \texttt{DITTO} not only mimics the watermark but also
exhibits an unexpected efficiency, mitigating the quality degradation that
can occur during the original watermarking process and generating more natural text.

\begin{figure}
    \centering
    \includegraphics[width=\columnwidth]{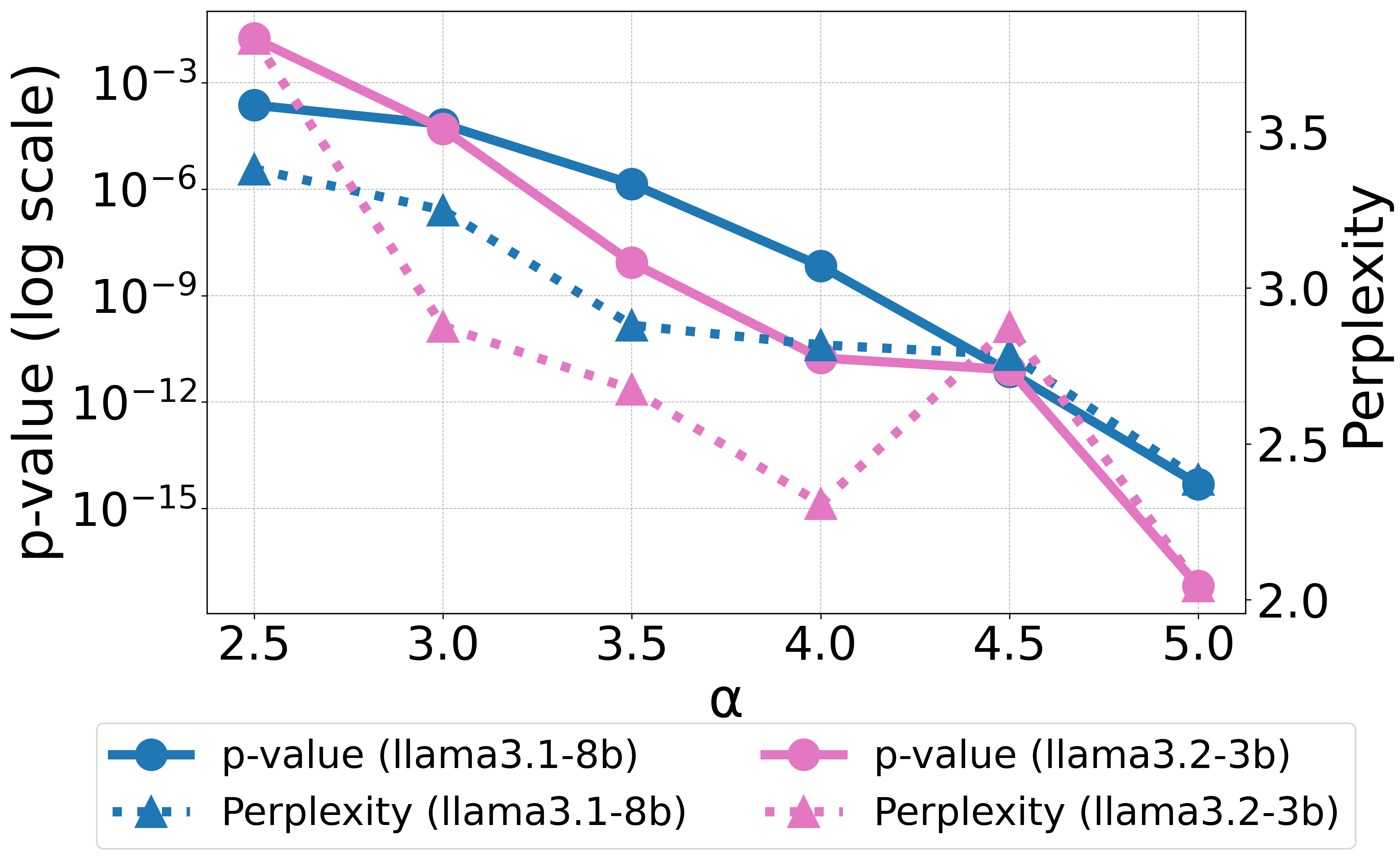}
    \caption{Impact of the scaling parameter $\alpha$ on spoofing performance against SynthID
    on the Dolly CW dataset.}
    \label{fig:synthid_alpha}
\end{figure}

Furthermore, consistent with our findings from the previous experiments (Figure~\ref{fig:alpha}),
we again analyze the trade-off between attack strength (\textit{i.e.}, $\alpha$)
and text quality (\textit{i.e.}, perplexity)
for spoofing sampling-based watermark.
Figure~\ref{fig:synthid_alpha} shows that as $\alpha$ increases, the p-value consistently
decreases while the perplexity fluctuates without a consistent upward trend.
Notably, and unlike our findings with KGW, the Llama3.1-8B model exhibits a more
stable perplexity pattern in the SynthID experiment.
This finding reinforces a key conclusion from Section~\ref{subsec:alpha_scaling}:
our attack's stealthiness appears to be an inherent trait, persisting even against SynthID
and strengthening our claim that its intensity can be increased without a clear cost to text quality.

\begin{table}[t]
    \centering\resizebox{\columnwidth}{!}{
    \small
    \begin{tabular}{l|ccccc}
        \toprule
        & \multicolumn{4}{c}{TPR@FPR=} 
        & \multirow{2}{*}{ppl $\downarrow$} \\
        size of $D_T$ &  10\%$\uparrow$ & 5\%$\uparrow$& 1\%$\uparrow$ & 0.1\%$\uparrow$ \\
        \midrule
        10k (\texttt{DITTO}) & \textbf{0.81} & \textbf{0.76} & \textbf{0.70} & 0.51 & 4.18 \\
        8k & 0.72 & 0.70 & 0.65 & \textbf{0.55} & 2.80 \\
        6k & 0.71 & 0.65 & 0.58 & 0.44 & 2.91   \\
        4k & 0.68 & 0.64 & 0.49 & 0.41 & 2.63 \\
        \bottomrule
    \end{tabular}}
    \caption{Impact of $D_T$ size on spoofing performance for the Dolly CW dataset.}
    \label{tab:train_scale}
\end{table}

Our analysis culminates in two key findings that provide compelling evidence for eneralizability:
\begin{enumerate}
    \item \textbf{Versatility across Schemes:} The attack is not confined to a single watermarking architecture, demonstrating high effectiveness against both green-list (KGW) and fundamentally different sampling-based (SynthID) methods.
    \item \textbf{Stealth without Compromise:} The attack breaks the conventional trade-off between strength and quality. Its intensity can be increased without producing a consistent, discernible degradation in text quality, making it highly evasive.
\end{enumerate}

We conclude that \texttt{DITTO} is not confined to a specific watermarking algorithm but possesses
a universal ability to learn and imitate artificial statistical bias.
% Moreover, the negligible cost to text quality during this process establishes
% it as a practical and formidable attack vector.

\begin{table*}[h!]
    \centering\resizebox{\textwidth}{!}{
    \small
    \begin{tabular}{l|ccccc|ccccc}
        \toprule
        & \multicolumn{5}{c|}{\textbf{MMW Bookreport}} & \multicolumn{5}{c}{\textbf{Dolly CW}} \\
        \cmidrule(lr){2-11}
        & \multicolumn{3}{c}{TPR@FPR=} 
        & \multirow{2}{*}[-2pt]{\shortstack{median\\p-value $\downarrow$}} & \multirow{2}{*}{ppl $\downarrow$} 
        & \multicolumn{3}{c}{TPR@FPR=} 
        & \multirow{2}{*}[-2pt]{\shortstack{median\\p-value $\downarrow$}} & \multirow{2}{*}{ppl $\downarrow$}  \\
        \cmidrule(lr){2-4} \cmidrule(lr){7-9}
         &  10\%$\uparrow$ & 1\%$\uparrow$ & 0.1\%$\uparrow$ & & & 10\%$\uparrow$ & 1\%$\uparrow$ & 0.1\%$\uparrow$ & & \\
        \midrule
        \multicolumn{11}{c}{\textbf{KGW}}  \\
        \midrule
        Llama3.2-3b \textit{(non-wm)} & 0.34 & 0.09 & 0.01 & 2.40E-01 & 3.00 & 0.33 & 0.14 & 0.09 & 3.20E-01 & 3.72 \\
        \quad + \textit{distill-only} & 0.57 & 0.34 & 0.19 & 6.08E-02 & \textbf{2.89} & 0.79 & 0.65 & 0.55 & 2.83E-04 & 4.61 \\
        \quad \textbf{+ \texttt{DITTO}}   & \textbf{0.81} & \textbf{0.70} & \textbf{0.51} & \textbf{8.00E-04} & 4.18 & \textbf{0.84} & \textbf{0.76} & \textbf{0.64} & \textbf{4.83E-05} & \textbf{3.28} \\
        \midrule
        Llama3.2-1b \textit{(non-wm)} & 0.30 & 0.08 & 0.03 & 2.83E-01 & 3.29 & 0.35 & 0.11 & 0.07 & 3.24E-01 & 8.39 \\
        \quad + \textit{distill-only} & 0.96 & 0.93 & 0.90 & 2.60E-09 & 2.65 & 0.56 & 0.44 & 0.41 & 4.65E-02 & 3.91 \\
        \quad \textbf{+ \texttt{DITTO}} & \textbf{0.99} & \textbf{0.99} & \textbf{0.97} & \textbf{5.48E-17} & \textbf{2.44} & \textbf{0.91}  & \textbf{0.86}  & \textbf{0.75}  & \textbf{1.78E-07}  & \textbf{2.95} \\
        \midrule
        \multicolumn{11}{c}{\textbf{SynthID}}  \\
        \midrule
        Llama3.2-3b \textit{(non-wm)} & 0.24 & 0.09 & 0.02  & 4.04E-01 & \textbf{3.00} & 0.14 & 0.06 & 0.02 & 6.11E-01  & 3.72  \\
        \quad + \textit{distill-only} & \textbf{0.39}  & 0.11 & 0.03 & \textbf{2.07E-01} & 3.01 & 0.62 & 0.43 & 0.36 & 2.26E-02 & 5.66 \\
        \quad \textbf{+ \texttt{DITTO}}   & 0.32 & \textbf{0.13}  & \textbf{0.08}  & 2.62E-01  & 3.81  & \textbf{0.88} & \textbf{0.84}  & \textbf{0.80}  & \textbf{7.10E-12}  & \textbf{2.79} \\
        \midrule
        Llama3.2-1b \textit{(non-wm)} & 0.21  & 0.07 & 0.03 & 4.29E-01 & \textbf{3.29} & 0.26 & 0.15 & 0.10 & 4.82E-01 & 8.39 \\
        \quad + \textit{distill-only} & \textbf{0.37} & \textbf{0.18} & \textbf{0.10} & \textbf{2.03E-01} & 5.56 & 0.50 & 0.38 & 0.32 & 9.68E-02 & 3.99 \\
        \quad \textbf{+ \texttt{DITTO}} & 0.21 & 0.06 & 0.01 & 3.63E-01 & 3.92 & \textbf{0.90} & \textbf{0.87} & \textbf{0.81} & \textbf{8.12E-12} & \textbf{2.88} \\
        \bottomrule
    \end{tabular}}
    \caption{Comparison of spoofing effectiveness between \texttt{DITTO} and the passive distillation \textit{(distill-only)}.}
    \label{tab:sft_only}
\end{table*}

\subsection{Potential of Synthetic Data}
\label{subsec:data_scaling}

The potency of \texttt{DITTO} is intrinsically linked to its utilization of synthetic data,
a resource that is theoretically inexhaustible as it can be generated directly from
the target teacher model. As demonstrated in Table~\ref{tab:train_scale},
the effectiveness of the spoofing attack generally follows a scaling law, where the
TPR across most FPR levels exhibits an overall upward trend as the size of the
training dataset~($D_T$) increases.
This positive correlation indicates that the extracted watermark signal
becomes increasingly authoritative as the student model is exposed to more
watermarked outputs.

\subsection{Why Distillation is Not Enough}
\label{subsec:sft_only}

The results in Table~\ref{tab:sft_only} underscore the critical necessity of a
framework that explicitly isolates and re-applies watermark signals, rather than
relying on passive inheritance. 
We compare \texttt{DITTO} against a \textit{distill-only} baseline, which refers to models
trained using only SFT without watermark extraction (Sec.~\ref{subsec:ews}).
While standard distillation allows a
student model to exhibit radioactive traces of a teacher's watermark, this approach
is insufficient for an effective spoofing attack.
As depicted in Table~\ref{tab:sft_only}, \texttt{DITTO} consistently outperforms the
\textit{distill-only} approach across nearly all metrics and model configurations.
These results demonstrate that relying solely on distillation fails to capture the
full intensity of the target watermark.
Therefore, the specialized \texttt{DITTO} framework is essential for achieving a 
high-fidelity spoofing attack that simple supervised fine-tuning cannot replicate.

\subsection{Other Victims}
\label{subsec:heterogeneous_teachers}

While previous experiments focused on scenarios where the victim and attacker
sharing the same model architecture, it is crucial to evaluate whether
\texttt{DITTO} remains effective against victims from heterogeneous model families.
To assess this cross-family generalizability, we conducted experiments using
GPT-OSS 20B and Phi-4 14B as target victims, while employing Llama 3.2 3B and 1B as
the attack models. As presented in Table~\ref{tab:heterogeneous_teachers},
\texttt{DITTO} successfully replicates the target watermark even when the attacker
and victim belong to entirely different model families.
% Although the spoofing intensity is slightly less pronounced compared to
% intra-family attacks,
The results consistently show significantly higher
watermark detection rates and lower median $p$-values than the non-watermarked
baselines, demonstrating that \texttt{DITTO} is not constrained by architectural
similarity.

\begin{table}[t]
    \centering\resizebox{\columnwidth}{!}{
    \small
    \begin{tabular}{l|cccc}
        \toprule
        & \multicolumn{3}{c}{TPR@FPR=} 
        & \multirow{2}{*}[-2pt]{\shortstack{median\\p-value $\downarrow$}} \\
        &  10\%$\uparrow$ & 1\%$\uparrow$ & 0.1\%$\uparrow$ \\
        \midrule
        3b \textit{(non-wm)} & 0.34 & 0.09 & 0.01 & 2.40E-01 \\
        \textbf{\texttt{DITTO} (3b)} $\rightarrow$ GPT & 0.60 &  0.50 & \textbf{0.42} & 1.18E-02 \\
        \textbf{\texttt{DITTO} (3b)} $\rightarrow$ Phi & \textbf{0.67} &  \textbf{0.55} & 0.35 & \textbf{7.06E-03} \\
        \midrule
        1b \textit{(non-wm)} & 0.30 &  0.08 & 0.03 & 2.83E-01 \\
        \textbf{\texttt{DITTO} (1b)} $\rightarrow$ GPT & 0.50 &  0.43 & 0.35 & 9.77E-02 \\
        \textbf{\texttt{DITTO} (1b)} $\rightarrow$ Phi & \textbf{0.65} &  \textbf{0.52} & \textbf{0.42} & \textbf{5.45E-03} \\
        \bottomrule
    \end{tabular}}
    \caption{Cross-family generalizability of the framework on the Dolly CW dataset. Arrows ($\rightarrow$) indicate the direction from the attacker model to the victim model.}
    \label{tab:heterogeneous_teachers}
\end{table}

\section{Conclusion}
\label{sec:conclusion}
% We challenge the prevailing premise that watermark detection constitutes 
% reliable proof of authorship and demonstrates that this trust can be 
% actively weaponized. 
% We introduce \texttt{DITTO}, a spoofing framework that repurposes watermark 
% radioactivity through knowledge distillation to learn and replay a target 
% model’s statistical fingerprint, 
% enabling seamless injection of that signature into a different model 
% and thereby supporting convincing misattribution at scale. 
% Empirically, the attack not only generalizes across various watermarking schemes,
% such as n-gram and sampling-based methods, 
% but also overcomes the inherent strength-quality trade-off 
% by achieving high spoofing intensity without a 
% corresponding loss in textual fluency.
% These findings reveal a systemic vulnerability with direct consequences 
% for provenance, moderation, licensing compliance, and incident response 
% in high-stakes settings. 
% The field should therefore pivot from presence testing to 
% authenticity verification, prioritizing adversarial resilience as a 
% design axiom and exploring mechanisms that bind a watermark to model identity, 
% for example cryptographic commitments.
\texttt{DITTO} repurposes watermark radioactivity to spoof LLM authorship across
diverse schemes without sacrificing text quality.
Our findings prove that watermark presence is no longer a reliable proof of origin,
necessitating a paradigm shift from mere presence detection toward robust
authenticity verification and cryptographic model-binding.

\section*{Limitations}
While this study provides a new perspective on LLM watermark security, it has several
limitations. First, the efficacy of \texttt{DITTO} is constrained by the fidelity of watermark
inheritance, which we found to be sensitive to model size and architectures. As our
primary focus was on demonstrating the attack’s feasibility, we did not perform an
exhaustive optimization of these variables. This leaves a significant opening for
future research to systematically map the boundary conditions of watermark
radioactivity. Second, this work is inherently adversarial and does not prescribe
specific defense mechanisms. By intentionally focusing on the vulnerability side, we
aim to provide a necessary baseline for the community. Investigating how to distort
or neutralize these leaked signals during the distillation process remains a crucial
next step for developing more resilient content protection systems.

\section*{Ethical Considerations}
The \texttt{DITTO} framework, by design, demonstrates a potent method for watermark spoofing that could be exploited for malicious purposes, such as spreading disinformation and falsely attributing it to trusted sources. However, our primary objective is not to provide a tool for malicious actors, but to proactively identify and expose a critical vulnerability in the current paradigm of LLM watermarking. By revealing the fragility of the core assumption that a detected watermark guarantees authorship, our work serves as a crucial red-teaming exercise for the AI safety community.
We argue that the responsible disclosure of this vulnerability is essential to next-generation watermarking techniques where adversarial resilience is treated as a fundamental design. Ultimately, our findings call for a paradigm shift from mere presence detection to developing technologies capable of distinguishing authentic watermarks from expertly imitated ones, thereby ensuring genuine authorship and fostering more secure AI ecosystems.

\section*{Acknowledgments}
This research was supported by the NRF grant~(RS-2025-00562134) and the AI Graduate School Program~(RS-2020-II201361) funded by the Korean government.


\bibliography{custom}

\appendix

\section{Detailed Results of the Spoofing Strength Analysis}
\label{appendix:full}

We provide the complete numerical data for the spoofing strength scaling experiment,
which is discussed in Section~\ref{subsec:alpha_scaling} and visualized in Figure~\ref{fig:alpha} of the main text.

Table~\ref{tab:alpha_full} details the median p-value and perplexity for each increment of the scaling parameter
$\alpha$ from 2.5 to 5.0. The results are presented for each experimental configuration, covering the two
imitation scenarios: 1) Llama3.2-3B imitates the watermark of Llama3.1-8B
and 2) Llama3.2-1B imitates the watermark of Llama3.2-3B
on the two datasets, MMW Bookreport and Dolly CW.

Consistent with the analysis in the main body, the data clearly shows a decrease in p-value as $\alpha$ increases,
indicating a stronger statistical significance for the spoofing attack.
Concurrently, the perplexity values fluctuate without a clear upward trend, providing the specific numerical evidence
that supports our conclusion regarding the lack of a severe trade-off between attack strength and text quality.

\begin{table*}[hbt!]
    \centering\resizebox{\textwidth}{!}{
    \small
    \begin{tabular}{l|ccccc|ccccc}
        \toprule
        & \multicolumn{5}{c|}{\textbf{MMW Bookreport}} & \multicolumn{5}{c}{\textbf{Dolly CW}} \\
        \cmidrule(lr){2-11}
        & \multicolumn{3}{c}{TPR@FPR=} 
        & \multirow{2}{*}[-2pt]{\shortstack{median\\p-value $\downarrow$}} & \multirow{2}{*}{perplexity $\downarrow$} 
        & \multicolumn{3}{c}{TPR@FPR=} 
        & \multirow{2}{*}[-2pt]{\shortstack{median\\p-value $\downarrow$}} & \multirow{2}{*}{perplexity $\downarrow$}  \\
        \cmidrule(lr){2-4} \cmidrule(lr){7-9}
         &  10\%$\uparrow$ & 1\%$\uparrow$ & 0.1\%$\uparrow$ & & & 10\%$\uparrow$ & 1\%$\uparrow$ & 0.1\%$\uparrow$ & & \\
        \midrule
        \multicolumn{11}{c}{\textbf{Llama3.1-8b (victim) mimicked by Llama3.2-3b (attacker)}} \\
        \midrule
        $\alpha=2.5$ &  0.62 & 0.26 & 0.08 & 6.83E-02 & \textbf{3.27} & 0.71 & 0.48 & 0.25 & 1.40E-02 & 3.86 \\
        $\alpha=3$   &  0.76 & 0.46 & 0.22 & 1.99E-02 & 3.65 & 0.79 & 0.57 & 0.45 & 4.36E-03 & 4.13 \\
        $\alpha=3.5$ &  0.79 & 0.56 & 0.38 & 4.36E-03 & 3.94 & 0.82 & 0.54 & 0.47 & 3.09E-03 & 4.31 \\
        $\alpha=4$   &  \textbf{0.87} & 0.65 & 0.46 & 1.83E-03 & 4.54 & 0.82 & 0.69 & 0.55 & 4.32E-04 & \textbf{3.23} \\
        $\alpha=4.5$ &  0.81 & \textbf{0.70} & \textbf{0.51} & \textbf{7.97E-04} & 4.18 & 0.84 & \textbf{0.76} & 0.64 & 4.83E-05 & 3.28 \\
        $\alpha=5$   &  0.73 & 0.61 & 0.49 & 1.83E-03 & 3.81 & \textbf{0.91} & \textbf{0.76} & \textbf{0.72} & \textbf{3.75E-05} & 3.25 \\
        \midrule
        \multicolumn{11}{c}{\textbf{Llama3.2-3b (victim) mimicked by Llama3.2-1b (attacker)}} \\
        \midrule
        $\alpha=2.5$ &  0.98 & 0.94 & 0.84 & 2.04E-06 & 3.79 & 0.72 & 0.63 & 0.46 & 2.34E-03 & 5.05 \\
        $\alpha=3$   &  \textbf{0.99} & 0.97 & 0.95 & 3.47E-10 & 3.57 & 0.83 & 0.74 & 0.64 & 2.05E-05 & 3.34 \\
        $\alpha=3.5$ &  0.97 & 0.95 & 0.93 & 3.07E-12 & 3.23 & \textbf{0.92} & 0.82 & 0.68 & 2.04E-06 & 4.33 \\
        $\alpha=4$   &  0.96 & 0.95 & 0.92 & 1.65E-17 & 2.81 & \textbf{0.92} & \textbf{0.89} & 0.79 & 1.74E-08 & 2.86 \\
        $\alpha=4.5$ &  \textbf{0.99} & \textbf{0.99} & \textbf{0.97} & 5.48E-17 & 2.44 & 0.91 & 0.86 & 0.75 & 1.78E-07 & 2.95 \\
        $\alpha=5$   &  0.94 & 0.91 & 0.89 & \textbf{7.00E-20} & \textbf{2.29} & \textbf{0.92} & 0.87 & \textbf{0.81} & \textbf{5.95E-10} & \textbf{2.80} \\
        \bottomrule
    \end{tabular}}
    \caption{Full results for the spoofing strength analysis, corresponding to the trends visualized in Figure~\ref{fig:alpha}.
    We report for two imitation scenarios across both datasets.}
    \label{tab:alpha_full}
\end{table*}

\section{Algorithms of \texttt{DITTO}}
We present the overall architecture and procedural details of \texttt{DITTO} through a series of formal algorithms.
The \texttt{DITTO} framework is composed of three core modules: (1) Watermark Inheritance, (2) Watermark Extraction and (3) the Spoofing Attack.
Algorithm~\ref{alg:inheritance} describes the inheritance process, where a student model learns the teacher's watermark patterns via knowledge distillation.
Next, Algorithm~\ref{alg:extraction} details the extraction process, where the inherited watermark is isolated as a quantitative signal (EWS).
Finally, Algorithm~\ref{alg:spoofing} shows the final attack stage, where this extracted signal is injected into an attack model to perform a
real-time spoof of the target watermark.
% kd
\begin{algorithm*}[hbt!]
\caption{Watermark Inheritance via Knowledge Distillation}
\label{alg:inheritance}
\begin{algorithmic}[1]
    \State \textbf{Input:} Watermarked teacher model $M_T$, original student model $M_O$, watermark strength $\delta$
    \State \textbf{Output:} Fine-tuned student model $M_S$ with inherited watermark
    \Statex 

    \Comment{Phase 1: Generate KGW-Watermarked Dataset}
    \State $D_T \gets \emptyset$
    \For{each prompt $p$ in a predefined set of prompts}
        \State $x \gets p$ \Comment{Initialize the sequence with a prompt}
        \While{not \textsc{EndOfSequence}(x)}
            \State $l \gets M_T(x_{1:t-1})$ \Comment{Get original logits from the teacher}
            \State $V_{\text{green}} \gets \textsc{GetGreenList}(x_{t-1})$ \Comment{Determine green list based on the previous token}
            \For{each token $v \in V_{\text{green}}$}
                \State $l_v \gets l_v + \delta$ \Comment{Add bias $\delta$ to logits of green tokens}
            \EndFor
            \State $x_t \sim \text{softmax}(l)$ \Comment{Sample next token from the modified distribution}
            \State $x \gets x \oplus x_t$ \Comment{Append the new token to the sequence}
        \EndWhile
        \State $D_T \gets D_T \cup \{x\}$
    \EndFor
    \Statex

    \Comment{Phase 2: Supervised Fine-Tuning of Student Model}
    \State Let $\theta_O$ be the initial parameters of $M_O$
    \State $\theta_S \gets \underset{\theta_O}{\text{arg max}} \sum_{x \in D_T} \sum_{i=1}^{|x|} \log P(x_i | x_{1:i-1}; \theta_O)$
    \State $M_S \gets M_O \text{ with parameters updated to } \theta_S$
    \Statex
    
    \State \Return $M_S$
\end{algorithmic}
\end{algorithm*}

% stealing
\begin{algorithm*}[hbt!]
\caption{Watermark Extraction}
\label{alg:extraction}
\begin{algorithmic}[1]
\State \textbf{Input:} Trained student model $M_S$, original student model $M_O$, training dataset $D_T$
\State \textbf{Output:} Extracted Watermark Signal (EWS) function

\State \Comment{Calculate the context-independent global bias}
\State $\delta_{\text{global}} \gets \mathbb{E}_{c \in D_T}[l_{M_S}(c)] - \mathbb{E}_{c \in D_T}[l_{M_O}(c)]$

\State \Comment{Calculate prefix-specific local biases for all n-gram prefixes $p$ in $D_T$}
\For{each unique prefix $p$ in $D_T$}
    \State $D_{T|p} \gets \{c \in D_T \mid c \text{ ends with } p\}$
    \State $\delta_p \gets \mathbb{E}_{c \in D_{T|p}}[l_{M_S}(c)] - \mathbb{E}_{c \in D_{T|p}}[l_{M_O}(c)]$
\EndFor

\State \Comment{Define the final EWS as a weighted sum of the biases}
\Function{EWS}{$c$}
    \State Let $\text{prefixes}(c)$ be the set of n-gram prefixes in context $c$
    \State $\text{EWS}(c) \gets \delta_{\text{global}} + \sum_{p \in \text{prefixes}(c)} w(p) \cdot \delta_p$
    \State \Comment{$w(p)$ is a weight based on prefix frequency}
    \State \textbf{return} $\text{EWS}(c)$
\EndFunction

\State \textbf{return} EWS function
\end{algorithmic}
\end{algorithm*}

% spoofing attack
\begin{algorithm*}[hbt!]
\caption{Spoofing Attack}
\label{alg:spoofing}
\begin{algorithmic}[1]
    \State \textbf{Input:} Attacker model $M_O$, EWS function, scaling parameter $\alpha$, initial context $c_0$
    \State \textbf{Output:} Text sequence $T$ with spoofed watermark
    \Statex

    \State $T \gets c_0$ \Comment{Initialize the generated text with a starting prompt}
    \State $c \gets c_0$ \Comment{Initialize the current context}
    \While{not \textsc{EndOfSequence()}}
        \State $x_{next} \gets \textsc{GenerateToken}(M_O, \text{EWS}, \alpha, c)$ \Comment{Generate the next token}
        \State $T \gets T \oplus x_{next}$ \Comment{Append the new token to the text sequence}
        \State $c \gets T$ \Comment{Update the context for the next step}
    \EndWhile
    \State \Return $T$
    \Statex

    \Procedure{GenerateToken}{$M_O$, EWS, $\alpha$, $c$}
        \State $l_{\text{original}} \gets M_O(c)$ \Comment{Get original logits from the attacker's model}
        \State $s_{\text{watermark}} \gets \text{EWS}(c)$ \Comment{Retrieve the extracted watermark signal}
        \State $l_{\text{modified}} \gets l_{\text{original}} + \alpha \cdot s_{\text{watermark}}$ \Comment{Inject the watermark signal}
        \State $P_{\text{modified}} \gets \text{softmax}(l_{\text{modified}})$ \Comment{Compute the new probability distribution}
        \State $x_{\text{sampled}} \gets \textsc{Sample}(P_{\text{modified}})$ \Comment{Sample from the modified distribution}
        \State \Return $x_{\text{sampled}}$
    \EndProcedure
\end{algorithmic}
\end{algorithm*}

\section{Dataset Examples}
Examples from the datasets utilized in our experiments are presented below.
The samples are drawn from the MMW (MarkMyWords) Bookreport benchmark and the Dolly-15k dataset, which served as our evaluation task.

The examples from Dolly-15k showcase the diversity of tasks our models were trained on, ranging from open-domain question answering and brainstorming
to information extraction from a given context. The MMW Bookreport sample, in contrast, exemplifies the long-form generation task used in our evaluation.

\begin{myrqbox}{Example of MMW Bookreport}
\textbf{"task\_type"}: "book\_report", \\
\textbf{"system\_message"}: "You are a helpful assistant. Always answer in the most accurate way.", \\
\textbf{"user\_message"}: "Write a book report about 'Pride and Prejudice', written by Jane Austen."
\end{myrqbox}

\begin{myrqbox}{Example of Dolly CW 1}
\textbf{"instruction"}: "Who is called Napoleon of Iran?", \\
\textbf{"context"}: "", \\
\textbf{"response"}: "Nadir Shah", \\
\textbf{"category"}: "open\_qa"
\end{myrqbox}

\begin{myrqbox}{Example of Dolly CW 2}
\textbf{"instruction"}: "Why capri is so famous", \\
\textbf{"context"}: "", \\
\textbf{"response"}: "Capri is famous Italian island for it's nice ocean color and it's mountain view", \\
\textbf{"category"}: "brainstorming"
\end{myrqbox}

\begin{myrqbox}{Example of Dolly CW 3}
\textbf{"instruction"}: "What is Github Copilot?", \\
\textbf{"context"}: "GitHub Copilot is a cloud-based artificial intelligence tool developed by GitHub and OpenAI to assist users of Visual Studio Code, Visual Studio, Neovim, and JetBrains integrated development environments (IDEs) by autocompleting code. Currently available by subscription to individual developers, the tool was first announced by GitHub on 29 June 2021, and works best for users coding in Python, JavaScript, TypeScript, Ruby, and Go.", \\
\textbf{"response"}: "GitHub Copilot is a cloud-based artificial intelligence tool developed by GitHub and OpenAI", \\
\textbf{"category"}: "closed\_qa"
\end{myrqbox}

\section{Examples of Spoofed Outputs}

We conduct a spoofing attack where a smaller model, Llama 3.2-3B, 
is trained to imitate the watermark of a larger model, 
Llama 3.1-8B, on the MMW Bookreport dataset.
% \begin{itemize}
%     \item You are a helpful assistant. Always answer in the most accurate way. Write a book report about `Great Expectations', written by Charles Dickens.
% \end{itemize}
\begin{mdframed}
    You are a helpful assistant. Always answer in the most accurate way. Write a book report about `Great Expectations', written by Charles Dickens.
\end{mdframed}
\begin{myrqbox}{Model answer}
The Great Expectations is a classic novel written by Charles Dickens, published in 1861. The story follows the journey of Pip, an orphan boy who is raised by his abusive sister, and her friend, an old man who is imprisoned for 20 years for murder. The story is an adaptation of the classic fairy tale `The Adventures of the Millionaire and the Genius, The King and the Genius and The Great and the Good'. The story is told through the eyes of the protagonist, Pip, and is written in the first-person narrative. The narrative is written in the past tense and is divided into three sections: the childhood, the present, and the future. The story is known for its themes of love, redemption, and the struggle between the old and the new. The characters of Miss Havisham, Abel Magwitch, and Miss Havisham's sister, Emily, are crucial to the story and are known for their unique personalities and motivations. The author uses vivid imagery
\end{myrqbox}

We conduct a spoofing attack where a smaller model, Llama 3.2-3B, 
is trained to imitate the watermark of a larger model, Llama 3.1-8B, 
on the Dolly CW dataset.
% \begin{itemize}
%     \item Question: Classify each as a chip or a cracker: Goldfish, Doritos, Cheez-It, Cheetos.
%     \item Context: None
% \end{itemize}

\begin{mdframed}
Question: Classify each as a chip or a cracker: Goldfish, Doritos, Cheez-It, Cheetos.

\noindent Context: None
\end{mdframed}

\begin{myrqbox}{Model answer}
- Goldfish: Chip \\- Doritos: Chip \\- Cheez-It: Cracker \\- Cheetos: Cracker \\The correct classification is based on the texture and the primary ingredients of the snack. The classification can be based on the primary ingredients and the texture.
\end{myrqbox}

We conduct a spoofing attack where a smaller model, Llama 3.2-1b, 
is trained to imitate the watermark of a larger model, Llama-8b, on the MMW Bookreport dataset.

\begin{mdframed}
You are a helpful assistant. Always answer in the most accurate way. Write a book report about `Moby Dick', written by Herman Melville.
\end{mdframed}

\begin{myrqbox}{Model answer}
``Moby Dick" is an epic novel written by Herman Melville, first published in 1851. The story revolves around the protagonist, Captain Ahab, a vengeful and obsessive whaling captain who is determined to hunt down and kill a white sperm whale named Moby Dick. The novel is an allegory for the human condition, exploring themes such as revenge, obsession, and the search for meaning in life. The story is set in the mid-19th century and is based on Melville's own experiences as a whaler. The plot of the novel follows the crew of the whaleship Pequod, which is captained by the enigmatic and experienced Ishmael. The crew is made up of various characters, each with their own distinct personality and backstory. There is Queequeg, a skilled harpooneer from a South Pacific island, who becomes Ishmael's close friend and confidant. The crew also includes Stubb, a young
\end{myrqbox}

We conduct a spoofing attack where a smaller model, Llama 3.2-1b, 
is trained to mimic the watermark of a larger model, 
Llama-8b, on the Dolly CW dataset.

\begin{mdframed}
    Question: Give me some ideas on where to travel for good sun and beaches.
    \noindent Context: None
\end{mdframed}

\begin{myrqbox}{Model answer}
The Caribbean, Hawaii, and Southeast Asia are popular destinations known for their beautiful beaches and tropical sun. The Maldives and Seychelles are also popular for their clear waters and white sand beaches. The Mediterranean region also has many beautiful beaches and is known for its pleasant weather. The US, Mexico, and the Caribbean are also popular for their sun and beaches. The best time to visit these destinations is during the summer months when the sun is at its strongest. The best time to visit the US, Mexico, and the Caribbean is from May to October when the temperatures are warm and the risk of hurricanes is low. The best time to visit the Caribbean and Southeast Asia is from late fall to early spring when the temperatures are cooler and the risk of hurricanes is low. The best time to visit the Mediterranean region is from late spring to early fall when the temperatures are warm and the risk of drought is high. The best time to visit the Maldives and Seychelles is from late fall to early spring
\end{myrqbox}

\end{document}